\documentclass[,aps,nofootinbib,twocolumn]{revtex4}
\usepackage{graphicx}
\usepackage{amsmath,amssymb}
\input{epsf}
\usepackage{epsf}

\def\CQG{{\it Class. Quantum Gravity} }

\def\JMP{{\it J. Math. Phys.} }
\def\JHEP{{\it JHEP} }

\def\JCP{{\it JCAP} }

\def\NP{{ Nucl. Phys.} }
\def\PL{{Phys. Lett.} }
\def\PR{{Phys. Rev.} }
\def\PRL{{Phys. Rev. Lett.} }

\def\be{\begin{equation}}
\def\ee{\end{equation}}
\def\bea{\begin{eqnarray}}
\def\eea{\end{eqnarray}}
\def\ham{{{\cal{H}}_{\phi}^{loop}}}
\def\F{{F_{j,l}(p)}}
\def\mom{{p_{\phi}}}

\newcommand{\sgn}{\mathop{\mathrm{sgn}}}

\begin{document}
\title{Tachyon Matter in Loop Quantum Cosmology}
\author{A.~A.~Sen\footnote{email: anjan.sen@vanderbilt.edu}}
\affiliation{Department of Physics and Astronomy, Vanderbilt University,
Nashville, TN  ~~37235}
\begin{abstract}
An analytical approach for studying the cosmological scenario with a homogeneous tachyon field within the framework of loop quantum gravity is developed. Our study is based on the semi-classical regime where space time can be approximated as a continuous manifold, but matter Hamiltonian gets non-perturbative quantum corrections. A formal correspondence between classical and loop quantum cosmology is also established. The Hamilton-Jacobi method for getting exact solutions is constructed and some exact power-law as well as bouncing solutions are presented.
\end{abstract}
\maketitle     

\section{Introduction}
One of the outstanding problems in theoretical physics today is to describe the quantum regime for the gravitational field. A fully consistent quantum theory of gravity is still not available. Currently two most promising and rigorous approaches are String theory \cite{string} and Loop Quantum Gravity\cite{lqg}. String theory provides an elegant prescription to unify the fundamental interactions and it also has a well defined perturbative expansion, finite order by order. But the no-perturbative regime in string theory is still not well understood. On the other hand, Loop Quantum Gravity (LQG) is a background independent, non-perturbative candidate for the quantum theory of gravity. It is a formulation for canonical quantization of general relativity in terms of Ashtekar's variable\cite{ashtekar}. Some important achievements of this theory are the discreet spectra for the geometrical operators like area and volume\cite{area}, existence of the well-defined operators for the matter Hamiltonian\cite{operator} and the derivation of the Bekenstein-Hawking entropy formula for black holes\cite{entropy}. During last couple of years, the methods of LQG have been applied in cosmological context and is known as Loop Quantum Cosmology (LQC)\cite{lqc}. In this paper, we work in the LQC framework in the context of spatially flat homogeneous and isotropic models. 
        
In LQC, the universe has three evolutionary periods. Initially, the evolution is governed essentially by the discreet quantum nature of the space time predicted by the LQG and is described by the difference equations\cite{discreet1, discreet2}. The key feature of this period is the removal of initial big bang singularity of the space time\cite{discreet1}. Subsequently, the universe enters a semi-classical period of evolution, where the evolution equations are continuous but modified due to the non-perturbative quantum effects\cite{semiclass}. Finally, we have the usual classical regime where the cosmological dynamics is governed by the usual Einstein's equations. The intermediate semi-classical phase is the most interesting phase as far as LQC is concerned. The interesting features of this phase are avoidance of many singularities in cosmological\cite{cosmo} and gravitational collapse scenarios\cite{collap}, setting the initial condition for the inflation\cite{inflation}, possible signatures in Cosmic Microwave Background Radiation (CMBR) spectrum\cite{cmb}, nonsingular cyclic models\cite{cyclic} and trans-Planckian modifications to the frequency dispersion relation\cite{dispersion}. It has been shown recently that non perturbative effect in the semi-classical regime, leads to non-singular bouncing and oscillatory behavior for the universe\cite{bounce}. Scaling solutions in LQC have also been constructed\cite{scaling}  and correspondence between string inspired and loop quantum cosmologies has been found\cite{corres,param}. Although, in most of these studies, the loop quantum effect has been incorporated through the modification in the matter Hamiltonian, recently Ashtekar {\it et al.} have made a complete analysis in  LQC incorporating all modifications in the matter as well as the gravitational part of the Hamiltonian \cite{full1,full2}. By solving the Hamiltonian constraint numerically, they have shown that the big bang is replaced by a big bounce.

Recently, an effective scalar field theory governed by a Lagrangian density with a non-canonical kinetic term 
($\mathcal{L}=-V(\phi)F(X)$, where $X = -{1\over{2}}\partial^{\mu}\phi\partial_{\mu}\phi$), has attracted considerable attention in cosmology.  Such a model can lead to a late time accelerated expansion and is called  ``k-essence'' \cite{kessence}. A scalar field with a non-canonical kinetic term has also been investigated for an early universe inflationary scenario and is termed ``k-inflation'' \cite{kinflation}. One example of such a field is tachyon matter governed by the Lagrangian density $\mathcal{L}_{tach} = -V(\phi)\sqrt{1-\partial^{\mu}\phi\partial_{\mu}\phi}$. As discussed by Padmanabhan and Roy Choudhury \cite{PR}, this is the generalization of the Lagrangian of a relativistic particle. The Hamiltonian structure of tachyonic matter given by the above Lagrangian is very similar to that of a  special relativistic particle governed $\mathcal{L}= -m\sqrt{1 - \dot{q}^2}$ where $m$ and $q$ are the mass and generalized coordinate of the particle. This tachyon field can naturally arise in open string theory\cite{sen} and can provide a rich gamut of possibilities in cosmological context\cite{taccosmo}. In this work, we explore the cosmological scenario with a tachyon matter in the context of loop quantum cosmology concentrating in the semi-classical regime. We shall not try to connect the $\mathcal{L}_{tach}$ to its string theory origin, rather we shall treat it just as a field with a nonstandard kinetic term.

In section 2, we describe the tachyon matter in standard cosmology in Hamiltonian approach. In section 3, we generalize it to LQC and show that tachyon matter in the semi-classical regime of LQC always results a super-accelerating period.
A formal correspondence between classical cosmology and LQC is also established. In section 3, we describe the Hamilton-Jacobi formulation to generate exact solution and provide two examples for a bouncing universe. The paper ends with concluding remarks in section 4.

\section{Tachyon in Standard Cosmology}

Assuming a spatially flat $3+1$ dimensional homogeneous and isotropic FRW background, the line element is given by
\begin{equation}
ds^2 = - dt^2 + a^{2}(t)(dx^2 + dy^2 + dz^2),
\end{equation}
where $a(t)$ is the scale factor. The Lagrangian for the tachyon field is given by \cite{sen}

\begin{equation}
L = -\sqrt{-g}V(\phi)\sqrt{1-\dot{\phi}^2}= -a^{3}V(\phi)\sqrt{1-\dot{\phi}^2},
\end{equation}
where we assume the tachyon field to be time dependent, $\phi = \phi(t)$, and $V(\phi)$ is the potential for the tachyon field. One can calculate the conjugate momentum for the tachyon field:
\begin{equation}  p_{\phi} = {\partial{L}\over{\partial{\dot{\phi}}}} = {a^{3}V(\phi)\dot{\phi}\over{\sqrt{1-\dot{\phi}^2}}}.
\end{equation}
Using the Legendre transformation, one can now write the Hamiltonian for the tachyon field as
\begin{equation}
{\cal{H}}_{\phi}(\phi,p_{\phi}) = a^{3}\sqrt{V^{2} + a^{-6}p_{\phi}^2}.
\end{equation}

\noindent The Hamiltonian  equations are now given by
\begin{eqnarray}
\dot{\phi} = {\partial{H_{\phi}}\over{\partial{p_{\phi}}}} = {a^{-3}p_{\phi}\over{\sqrt{V^{2} + a^{-6}p_{\phi}^2}}}\nonumber\\
\dot{p_{\phi}} = -{\partial{H_{\phi}}\over{\partial{\phi}}} = -{a^{3}VV^{'}\over{\sqrt{V^{2} + a^{-6}p_{\phi}^2}}},
\end{eqnarray}
where prime denotes differentiation w.r.t $\phi$. One can use these two equations to get the standard equation of motion for the tachyon field as 
\begin{equation}
\ddot{\phi} = -(1-\dot{\phi}^{2})\left[3H\dot{\phi} +{V^{'}\over{V}}\right],
\end{equation}
where $H={\dot{a}\over{a}}$ is the Hubble parameter. One can also define a volume parameter $p^{3/2} = a^{3}$ and write the Hamiltonian for the tachyon field as
\begin{equation}
{\cal{H}}_{\phi}(\phi,p_{\phi}) = p^{3/2}\sqrt{V^{2} + p^{-3}p_{\phi}^2}
\end{equation}

The standard procedure for calculating the energy density and pressure for any field is to vary the action for the field with respect to the space time metric to obtain the energy-momentum tensor $T^{\mu}_{\nu}$ and then identify the energy density and pressure as different components of $T^{\mu}_{\nu}$. In the Hamiltonian formulation, although one does not have such standard method, one can still define the energy density and pressure in terms of the Hamiltonian\cite{golam}:

\begin{eqnarray}
\rho_{\phi} &=& p^{-3/2}{\cal{H}}_{\phi}\nonumber\\
P_{\phi} &=& -p^{-3/2}\left({2p\over{3}}{\partial{\cal{H_{\phi}}}\over{\partial{p}}}\right),
\end{eqnarray}
which satisfy the matter conservation equation $\dot{\rho_{\phi}} = - 3H(\rho_{\phi}+P_{\phi})$. Using equations (3) and (7), one can check that equation (8) yields the standard expressions for energy density and pressure for the tachyon field:

\begin{eqnarray}
\rho_{\phi} = \frac{V}{\sqrt{1-\dot{\phi}^2}}\nonumber\\
P_{\phi} = -V{\sqrt{1-\dot{\phi}^2}}
\end{eqnarray}

\section{Tachyon in Loop Quantum Cosmology}

In loop quantum cosmology the variable $p$ defined in the previous section, is known as redefined densitized triad and one of the basic phase space variables. The geometrical property of the space is contained in this variable. In loop quantum cosmology, the term $p^{-1}$ associated with the momentum operator $p_{\phi}$ in the effective scalar Hamiltonian is replaced by the eigenvalue $F_{j,l}(p)$ of the inverse densitized triad operator \cite{golam2} and one now writes the effective scalar Hamiltonian as
\begin{equation} 
{\cal{H}}_{\phi}^{loop} = p^{3/2}\sqrt{V^{2} + |F_{j,l}|^{3}p_{\phi}^{2}}.
\end{equation}
Here $(j,l)$ are two quantization parameters \cite{bojowald}. The parameter $j$ must take half integer values but otherwise arbitrary. It is related with the dimension of the representation while writing holonomy as a multiplicative operator. The parameter $l$ can take only values between zero and one and corresponds to different equivalent ways of writing the inverse power of the densitized traid  operator in terms of the poisson bracket of the basic variables. The eigenvalue of the inverse densitized triad operator, $\F$ is given by $F_{j,l}(p) = {p_{j}}^{-1}F_{l}(p/p_j)$ where $p_{j} = (1/3)\gamma j l_{p}^2 = {a_{i}^2}j/3 = a_{*}^2$ \cite{golam}. Here $\gamma = 0.13$ is the Barbero-Immirzi (B-I) parameter and $l_{pl}$ is the Planck length. Here $a_{i} = \sqrt{\gamma}l_{p}$ is the scale above which we assume classical continuous space time but below which the space time is discreet. The second scale is $a_{*}$ below which the modification to the geometrical density due to quantum effect is important and one gets the modified Hamiltonian for the matter field. For $j>3$, $a_{i} < a_{*}$, and there is a semi-classical regime $a_{i} < a << a_{*}$, where the space time can be considered as continuous but the quantum effects in the geometrical density is still important \cite{bojowald2}. 

There is another modification which is essentialy due to the discreet quantum geometric nature of the spacetime, as predicted by LQC in the regime $ a < a_{i}$. It leads to a $\rho^{2}$ modification of the FRW (Friedman-Robertson-Walker) equation of the form $H^{2} \propto \rho(1-\rho/\rho_{c})$ \cite{param},  where $\rho_{c} = 3/{\alpha \kappa \gamma^{2} l_{p}^2}$. Here $\alpha$ is a constant of the order of unity deteremined by the eigenvalue of the area operator, $\kappa = 8\pi G$.   This is similar to the modification in the FRW equation arising in the Randall-Sundrum brane world scenario \cite{RS} with a difference that the correction term $\rho^{2}$ comes with a negative sign leading to a nonsingular bouncing cosmology\cite{full1}. This modification in FRW equation is not only important in the highly energetic early universe when $\rho$ is comparable to $\rho_{c}$ which is of the order of Planck density but also in the future if the universe is dominated by phantom dark energy \cite{cald}. In a recent paper, Sami {\it et al.} have investigated the fate  of future singularities in a phantom dominated universe in the effective dynamics of LQC with a $\rho^{2}$ corrections in the FRW equation \cite{sami}. Also non-singular bouncing models in LQC with a $\rho^{2}$ modification in the FRW equation has been studied recently by Singh {\it et al.} \cite{param1}.

The domain in which the inverse scale factor modifications in the matter Hamiltonian given by eqn (10) is important, is solely determined by the parameter $j$ which determines $a_{*}$ whereas the domain in which the $\rho^{2}$ correction due to the discreet quantum effects is important depends on the value of the energy density. With an initial choice of matter configuration such that $\rho < \rho_{c}$, one can ignore the $\rho^{2}$ correction in the FRW equation and concentrate solely in the semi-classical regime $a_{i} < a << a_{*}$ with a modified matter Hamiltonian as in eqn (10). We will take this approach in our following study.

Using equation (10) for the effective Hamiltonian for the tachyon field,  one can now define the momentum of the tachyon field as 

\be
\mom = {\dot{\phi}V\over{|\F|^{3/2}\sqrt{p^{3}|\F|^{3} - \dot{\phi}^2}}},
\ee
which follows from the Hamiltonian equation of motion $\dot{\phi} = {\partial{\ham}\over{\partial{\mom}}}$. We now determine the modification in the energy density expression for the tachyon field. Classically, once we know the Hamiltonian ${\cal H}_{\phi}$, it is straighforward to calculate the energy density as in equation (8). But there are two ways to obtain energy density in the semi-classical regime of LQC. One way is to obtain the density operator $\rho_{q}={\cal H}/a^3$ and then take the semi-classical limit. Another way is to define the energy density as the ratio of modified Hamitonian to the volume. This is same as keeping the definition (8), only replacing the Hamiltonian ${\cal H}_{\phi}$ by the effective Hamiltonian $\ham$ \cite{semiclass}. It should be noted that while the first definition incorporates modifications both in energy and geometric energy eigenvalues, the second incorporates only the modification in the energy eigenvalue and does not receive any contribution from the modification in the behaviour of volume operator. In our subsequent calculations, we adopt the second type of convention for defining the energy density. Although the differences in the type of definitions do not affect the dynamical trajectories, however they do lead different effective Hubble rates and equation of states in semi-classical regime $a < a_{*}$ (for a detail study on how different ways of defining energy density may yield different results, see \cite{dispersion}).

The expressions for energy density and pressure for the tachyon field now becomes
 
\begin{eqnarray}
\rho_{\phi}^{loop} &=& {V p^{3/2} |\F|^{3/2}\over{\sqrt{p^{3}|\F|^3-\dot{\phi}^2}}}\nonumber\\
P_{\phi}^{loop} &=& -{V p^{3/2} |\F|^{3/2}\over{\sqrt{p^{3}|\F|^3-\dot{\phi}^2}}}\left[1+{\dot{\phi}^2 |\F|^{'}\over{p^{2}|\F|^4}}\right].
\end{eqnarray}

\noindent In the above equations, the prime denotes the differentiation w.r.t the argument $p$. One can also write the equation of state for the tachyon field,

\be
\omega_{\phi}^{loop} = -\left[1+{\dot{\phi}^2|\F|^{'}\over{p^{2}|\F|^4}}\right].
\ee

\noindent The function $F_{l}(p/p_{j})$ is given by

\be
F_l(q) = [ \frac{3}{2(l+2)(l+1)l}( 
(l+1) \left\{ (q + 1)^{l+2} -  |q - 1|^{l+2} \right\} ~-~ \nonumber
\ee
\be
(l+2) q \left\{ (q + 1)^{l+1} - \sgn(q - 1) |q - 1|^{l+1} \right\}]^{\frac{1}{1-l}}.
\ee

\vspace{5mm}
\noindent
One can recover the classical behavior $F_{l}(p/p_{j}) \rightarrow p_{j}/p$ for $a >> a_{*}$. In the semi-classical regime ($a_{i} < a << a_{*}$), $F_{l}(p/p_{j})$ is approximated as 

\be
F_{l}(p/p_{j}) = \left[{3(p/p_{j})\over{l+1}}\right]^{1\over{1-l}}.
\ee
Using equation (12) and (15), one can now write the relevant equations for the tachyon field in the semi-classical regime of the loop quantum gravity:

\be
3H^{2} = {V\over{\sqrt{1-A^{-1}(a_{*}/a)^{\alpha}\dot{\phi}^{2}}}} = \rho_{\phi}^{loop}
\ee
\be
2\dot{H} + 3H^{2} = {V\over{\sqrt{1-A^{-1}(a_{*}/a)^{\alpha}\dot{\phi}^{2}}}}[1+\nonumber
\ee
\be
{1\over{1-l}}A^{-1}(a_{*}/a)^{\alpha}\dot{\phi}^{2}] = - P_{\phi}^{loop}
\ee
\be
\ddot{\phi} - {3\dot{\phi}H\over{1-l}}\left[1 - A^{-1}(a_{*}/a)^{\alpha}\dot{\phi}^{2} + 2(2-l)\right] + \nonumber
\ee
\be
\left[A(a/a_{*})^{\alpha} - \dot{\phi}^{2}\right]{V^{'}\over{V}} = 0,
\ee
where prime denotes differentiation w.r.t $\phi$. The first equation in the above list is the Hamiltonian constraint equation, the second is the RayChaudhury equation and third one is the modified equation of motion for the tachyon field $\phi$. $A^{-1}$ in the above equations is given by $A^{-1} = [(l+1)/3]^{3/(1-l)}$ and $\alpha = 6(2-l)/(1-l)$ and is always greater than $6$ as $0 < l < 1$. It is easy to check that for $l = 2$, the equations (16)-(18) are identical to that for tachyon matter in usual classical cosmology. Using equations (16)-(18), one can now write

\be
2\dot{H} = {3H^{2}\over{1-l}}A^{-1}(a_{*}/a)^{\alpha}\dot{\phi}^{2}.
\ee
Hence $\dot{H}$ is always greater than zero for any value of $l$ in the allowed range $0 < l < 1$, showing that there is always a period of super-acceleration with tachyon field for loop quantum gravity in the semi-classical regime. This is unlike the tachyon field in the standard cosmology where $\dot{H}$ is always less than zero ruling out any super-accelerating regime. 

In the above set of equations (16)-(18), only two equations are independent as the third equation can be obtained from the other two. But as we have three unknowns, one has to assume some form for the potential $V(\phi)$ or for the scale factor $a(t)$. One can now introduce a new set of variables:

\bea
y = a^{-(1/(1-l))}, \hspace{5mm} \dot{\chi} = (1/A)({a_{*}\over{a}})^{\alpha/2}\dot{\phi},\nonumber\\
W(\chi(\phi)) = V(\phi)/(1-l)^2.
\eea

\noindent This transforms the set of equations (16) and (17) to

\be
3 F^{2}(t) = {W(\chi)\over{\sqrt{1-\dot{\chi}^2}}}
\ee
\be
2\dot{F(t)} = -{W(\chi)\dot{\chi}^{2}\over{\sqrt{1-\dot{\chi}^2}}}
\ee
where $F(t) = \dot{y}/y = -H/(1-l)$. The parameters $y$ and $F$ can be viewed as the rescaled scale factor and the Hubble parameter. The above set of equations are exactly same as the classical Einstein's equations with a tachyon field $\chi$ and potential $W(\chi)$. This formal correspondence between the classical and semi-classical loop quantum cosmology can have interesting consequences. As $0 < l < 1$, it relates a classical expanding model to a contracting loop quantum one and vice versa. As an example, in classical cosmology, Padmanabhan has earlier found a power-law model driven by an inverse square potential\cite{paddytac}:

\be
y(t) \sim t^{n}, \hspace{3mm} \chi = \chi_{0} + \sqrt{2\over{3n}} t, \hspace{3mm} W(\chi) = W_{0} (\chi-\chi_{0})^{-2}
\ee
where $\chi_{0}$ $W_{0}$ and $V_{0}$ are constants. Employing  the mapping (21), one can now write the corresponding solution in LQC as

\be
a(t) \sim t^{-m},\hspace{3mm} \phi = \phi_{0} + B t^{c}, \hspace{2mm} V(\phi) = V_{0}(\phi-\phi_{0})^{-p}
\ee

\noindent where $\phi_{0}$, $B$, $c$ and $V_{0}$ are constants, $m = n(1-l)$ and $p= 4/(2-m\alpha)$ . For $n >0$, it transforms a expanding classical solution to a contracting model in LQC whereas for $ n<0$ it does the opposite. It is interesting that although for power-law cosmology, the potential $W(\chi)$ in classical case is always inverse square function of the tachyon field $\chi$, in LQC, the power $p$ in the potential  $V(\phi)$ depends on the power $m$ of the scale factor $a(t)$.
 
\section{Hamilton-Jacobi Formulation with Tachyon Field}
The Hamilton-Jacobi formulation \cite{salopek} is a effective way of rewriting the equations of motion. It results an easier derivation of many cosmological solutions involving scalar fields  specially in case of inflation \cite{HJ}. It allows one to consider Hubble parameter as a fundamental quantity rather than potential $V(\phi)$. Once $H(\phi)$ is specified, one can directly get the form of the potential. The formalism has applications to general inhomogeneous cosmology, though we concentrate here on its homogeneous version.

If the Tachyon field $\phi(t)$ is a monotonically varying function of time, one can transform the equation (16) into the Hamilton-Jacobi Form:

\bea
V^{2}(\phi) &=& 9H^{4}\left[1-{(1-l)^{2}a^{\alpha}\over{\beta}}J_{\phi}\right]\nonumber\\
J_{\phi} &=& {\beta\over{(1-l)}}{\dot{\phi}\over{a^{\alpha}}},
\eea
where $\beta = a_{*}^{\alpha}/A$. One can now write the scale factor as

\be
a^{\alpha} = a_{0} + {\alpha \beta\over{1-l}}\int {H\over{J_{\phi}}} d\phi,
\ee
where $a_{0}$ is a constant of integration. Also by integrating the second equation in (25), one gets the time dependence of the tachyon field $\phi(t)$:

\be
t = {\beta\over{(1-l)}}\int {d\phi\over{J_{\phi}a^{\alpha}}}.
\ee
Also from equation (17), one can write 

\be
J_{\phi} = {2\over{3 H^{2}}}{dH\over{d\phi}}.
\ee 

\noindent Hence once we specify the Hubble parameter $H$ as a function of the tachyon field $\phi$, we can completely determine dynamics of the tachyon field. Here we show two examples of a bouncing model assuming specific dependence of Hubble parameter $H$ on $\phi$. We begin by considering a Hubble parameter of the form

\be
H = H_{1}\exp[g\phi]
\ee
where $H_{1}$ and $g$ are two arbitrary constants. Integrating equation (22), we now get 

\be
a^{\alpha} = a_{0} + {3\alpha\beta H_{1}\over{4g^{2}(1-l)}}\exp[2g\phi]
\ee

\noindent Assuming $a_{0} = {3\alpha\beta H_{1}\over{4g^{2}(1-l)}}$ without any loss of generality, one can now write the solution as:

\bea
\exp[g\phi] &=& Tan({\alpha\over{2}} t)\nonumber\\
V(\phi) &=& 3H_{1}^2 \exp[2g\phi](V_{1} - V_{2}\exp[-g\phi])^{1/2}\\
a^{\alpha}(t) &=& a_{0}Sec^{2}({\alpha\over{2}}t)\nonumber
\eea

\noindent where $V_{1} = (1-{\alpha (1-l)\over{2g}})$ and $V_{2} = {\alpha (1-l)\over{2g}}$. Here we implicitly assumed that $-\pi/2 < t < \pi/2$. The second class of bouncing model is given by,

\bea
H &=& {2\over{\alpha}}{\exp[{g\phi}]\over{\sqrt{(1+\exp[{2g\phi}])}}}\nonumber\\
\exp[g\phi] &=& Sinh[t]\nonumber\\
V(\phi) &=& V_{0}\exp[2g\phi]{(1-V_{1}\exp[2g\phi])^{1/2}\over{\exp[2g\phi]+1}}\\
a^{\alpha}(t) &=& a_{0}Cosh[2t]\nonumber
\eea
where $V_{0}$ is a constant, $g = \sqrt{(1-l)\alpha\over{3\beta}}$, $V_{1} = \alpha (1-l)/3$ and $a_{0} = {3\beta\over{g^{2}\alpha (1-l)}}$. Here the universe, after reaching a minimum radius at $t=0$,  bounces to an asymptotic de-Sitter phase in the late times.

\section{Conclusion}
In this paper, we have considered tachyon matter within the context of semi-classical loop quantum cosmology in the regime where the matter Hamiltonian receives a non-perturbative quantum correction. In this regime, the eigenvalue of the inverse volume operator has a power-law dependence on the scale factor. This is the first analytic approach in the semi-classical loop quantum cosmology involving a scalar field with a non canonical kinetic term (also known as ``k-essence'').

We have calculated the energy density and pressure for the tachyon matter in this effective semi-classical theory and have constructed the system of equations for the cosmic dynamics. One of the interesting features of tachyon matter in LQC is that it always results super-accelearting period during the semi-classical regime irrespective of the value of the model parameters. Hence tachyon matter in the semi-classical regime of the LQC always behaves as a phantom matter without having a wrong sign in front of its kinetic term in the action.

We have also established a formal correspondence between the Einstein's equations in LQC and those in classical cosmology. It relates an expanding universe in classical cosmology to a contracting one in LQC. A similar correspondence has also been found earlier by Lidsey with a minimally coupled scalar field with usual canonical kinetic term\cite{scaling}. We have also shown that, for a power-law cosmology ($a\sim t^{n}$), the potential $V(\phi)$ is of the form $V(\phi) \sim (\phi-\phi_{0})^{p}$, with $p$ depending on $n$. This is unlike the classical behavior, where for power-law cosmology, the potential for the tachyon is always inverse square function of the field $\phi$ irrespective of the power in the scale factor \cite{paddytac}.

Finally, we present the Hamilton-Jacobi method for obtaining exact solution of the system of equations in LQC once one chooses some specific form for the Hubble parameter $H$ as a function of field $\phi$. We present two examples of nonsingular bouncing solutions for the cosmic dynamics. one of them bounces to an asymptotically de-Sitter universe. These solutions can be useful for studying the evolution of perturbations in a nonsingular bouncy tachyon dominated universe.

This work is a first step towards studying k-essence field in the context of LQC and it will be interesting to investigate more general form for the Lagrangian of the k-essence field in LQC. Also, in our study, we have chosen an initial matter configuartion with $\rho < \rho_{c}$, thus neglecting the $\rho^{2}$ correction in the FRW equation coming from the discrete quantum nature of the spacetime. It will be really interesting to also consider such correction together with the modification in the matter Hamiltonian and do similar study. It has been earlier shown that scaling solutions in LQC with a $\rho^2$ correction in the FRW equation has dual relationship with those in Randall-Sundrum cosmology \cite{param}. Considering LQG corrections both in the gravitational and matter part of the Einstein's equations, it will be really interesting to study such duality in Tachyon cosmology.  It may be also interesting to consider such corrections in case of Tachyon inflation which although faces serious difficulty in standard cosmology \cite{kofman}, subsequent investigations have shown that $\rho^{2}$ corrections in FRW equation coming from brane models can alleviate the problem \cite{bento}. This will be our aim in future investigations.

\end{document}